\documentclass[a4paper,aps, 10pt,superscriptaddress,floatfix,nofooti nbib,notitlepage]{revtex4-1}
\usepackage[utf8]{inputenc}
\usepackage[normalem]{ulem}
\usepackage{amssymb}
\usepackage{graphicx,epsfig}
\usepackage{color}
\usepackage{numprint}
\usepackage{booktabs}
\usepackage{hhline}
\usepackage{bm, amsmath, amsfonts}
\usepackage{enumitem}
\usepackage{mathtools}
\usepackage{hyperref}
\usepackage{cleveref}
\usepackage{nccmath}
\usepackage{float}
\usepackage{mwe}
\hypersetup{
	colorlinks=true, 	
	linkcolor=blue,		
    citecolor=cyan,		
}

\usepackage[normalem]{ulem} 
\usepackage{multirow}

\newcommand{\Sa}{\mathbf{S}_1}
\newcommand{\Sb}{\mathbf{S}_2}
\newcommand{\Si}{\mathbf{S}_i}

\newcommand{\vek}[1]{\mathbf{#1}}
\newcommand{\tpre}{t_{pre}}
\newcommand{\trr}{t_{rr}}
\def\nn{\nonumber}

\newcommand{\IUCAA}{Inter-University Centre for Astronomy and
  Astrophysics, Post Bag 4, Ganeshkhind, Pune 411 007, India}

\newcommand{\WSU}{Department of Physics \& Astronomy, Washington State University, 1245 Webster, Pullman, WA 99164-2814, U.S.A.}

\newcommand{\Pennstate}{Department of Physics, The Pennsylvania State University, University Park, PA 16802, USA}

\newcommand{\IGC}{Institute for Gravitation \& the Cosmos, The Pennsylvania State University, University Park, PA 16802, USA}

\newcommand{\IITK}{Department of Physics, Indian Institute of Technology, Kanpur 208016, India}

\newcommand*{\rom}[1]{\expandafter\@slowromancap\romannumeral #1@}
\newcommand{\RNum}[1]{\uppercase\expandafter{\romannumeral #1\relax}}
\allowdisplaybreaks[1]

\begin{document}


\title{Effect of orbital eccentricity on the dynamics of precessing compact binaries}

\author{Khun Sang Phukon}
\email{khunsang@iitk.ac.in}
\affiliation{\IITK}

\author{Anuradha Gupta}
\email{axg645@psu.edu}
\affiliation{\Pennstate}
\affiliation{\IGC}

\author{Sukanta Bose}
\email{sukanta@iucaa.in}
\affiliation{\IUCAA}
\affiliation{\WSU}


\author{Pankaj Jain}
\email{pkjain@iitk.ac.in}
\affiliation{\IITK}

\begin{abstract}
We study precession dynamics of generic binary black holes in eccentric orbits using an effective potential based formalism derived in [M. Kesden et al., PRL {\bf 114}, 081103 (2015)].
This effective potential is used to classify binary black holes into three mutually exclusive spin morphologies. During the inspiral phase, binaries make transitions from one morphology to others. 
We evolve a population of binary black holes from an initial separation of $1000\mathbf{M}$ to a final separation of $10\mathbf{M}$ using post-Newtonian accurate evolution equations. We find that, given suitable initial conditions, a binary's eccentricity can follow one of three distinct evolutionary patterns: (i) eccentricity monotonically increasing until final separation, (ii) eccentricity rising after decaying to a minimum value, and (iii) eccentricity monotonically decreasing throughout the inspiral. The monotonic growth or growth after reaching a certain minimum of eccentricity is due to the effect of 2PN spin-spin coupling. Further, we investigate the morphology transitions in eccentric binaries and find that the probability of such binaries transiting from one to other is similar to those in circular orbits, implying that eccentricity plays a sub-dominant role in spin morphology evolution of a precessing binary black hole. We, hence, argue that the morphological classification of spin precession dynamics is a robust tool to constrain the formation channels of binaries with arbitrary eccentricity as well.

\end{abstract}

\maketitle

\section{Introduction} 
\label{sec:intro}

The recent detections of gravitational wave (GW) from the mergers of stellar mass binary black holes (BBHs) and binary neutron stars~\cite{Abbott:2016blz, Abbott:2016nmj, Abbott:2017gyy,gw170817,gw170814, Abbott:20170104} by advanced LIGO (aLIGO) and advanced Virgo detectors~\cite{advLigo2015,advVirgo2015} not only provide direct evidences of the existence of GWs but also open a new window of exploration to the Universe. These momentous discoveries have also confirmed the existence of BBHs in the cosmos. In the coming years, the networks of both current and planned detectors  will continue to resolve
 the population of compact binary coalescences particularly BBHs, providing a unique opportunity to understand astrophysics and relativity in strong gravity regime.

 At present, the formation scenarios of BBHs and the evolutionary processes of their progenitors are highly uncertain.  The properties of an ensemble of BBHs by means of the measurement of their merger rates, masses, eccentricities, spins, and redshift distributions can furnish crucial astrophysical information about the formation mechanisms of BBHs. Amongst these observables,  black hole (BH) spins provide the most promising means to constrain their formation channels~\cite{Mandel:2009nx, Belczynski:2001uc,Rodriguez:2015oxa,Barrett:2017fcw,Lower:2018seu,Breivik:2016ddj,Rodriguez:2016vmx,Gerosa:2017kvu,Vitale:2015tea,Talbot:2017yur, Farr:2017gtv}. Different formation mechanisms leave distinct imprints on BH spins at the time of binary formation. For example, in the dynamical formation scenario \cite{Gultekin:2004pm,OLeary:2005vqo,Grindlay:2005ym,Sadowski:2007dz,Ivanova:2007bu,Downing:2009ag,Miller:2008yw}, the BH spins are expected to be isotropically distributed with respect to the orbital angular momentum \cite{Rodriguez:2016vmx, Mandel:2009nx} while the spins in field model \cite{Belczynski:2001uc,Belczynski:2005mr} are mostly aligned with the orbital angular momentum \cite{Belczynski:2017gds}.
 The BH spin magnitude remains conserved while the orientations of BH spins can be significantly distorted during the inspiral due to spin-orbit and spin-spin interactions. 
 The spin-orbit misalignment produces relativistic precession of the spin ($\vek S_1, \vek S_2$) and orbital angular momentum ($\vek L$) about the total angular momentum  ($\vek J$) of the BBH. As a result, modulations in the amplitude and phase of GW signal are observed at the detector~\cite{Apostolatos:1994mx}. Measurement of spin orientations in GW signals can provide insights to the astrophysical processes that misalign/align the spins of BBHs during their formation, opening avenues to explore the interplay between astrophysics and relativity~\cite{Gerosa:2013laa,Gerosa:2018wbw}.
 
As GW detectors are preparing for the observation of a population of BBHs in the coming years, it is desirable to acquire a detailed understanding of the dynamics of precessing BBHs to maximize the scientific output of GW detectors' data.  The dynamics of precessing BBHs during inspiral, until BHs get sufficiently close to each other, can be described very accurately by post-Newtonian (PN) approximation to general relativity that provides a foundation to calculate the gravitational waveforms as well as the orbital evolution of compact binaries under radiative loss~\cite{Kidder:1995zr,Racine:2008qv}. Precession cause changes in the orientations of spins and orbital angular momentum of BBHs during their evolution on the precession time ($t_{pre}$). For spinning BBHs in circular orbits, $t_{pre}$ falls between the orbital time ($t_{orb}$) and radiation-reaction time ($t_{rr}$). This timescale hierarchy of precessing BBHs in circular orbits, which can be mathematically stated as $ t_{orb} < t_{pre} < t_{rr}$, has been extensively used for studying various aspects of binary orbital evolution in the literature. For example, Ref.~\cite{Schnittman:2004vq} used the inequality $t_{orb} < t_{pre}$ to solve the orbit-averaged PN spin precession equations, where  a set of equilibrium  configurations of spins and orbital angular momenta was discovered. These configurations are termed as spin-orbit resonances (SOR). For BBHs in these configurations, the three angular momenta, $\vek{L},\, \Sa$ and $\Sb$, remain coplanar with their relative orientations slowly varying throughout the inspiral. 
The SOR configurations are useful in describing PN spin dynamics of precessing BBHs~\cite{Gupta:2013mea,Kesden:2010yp}. 
These solutions have been used in the studies of precession dynamics to predict the spin distribution of a population of BBHs before their merger, and the distribution of final spins of the daughter BHs formed after merger~\cite{Kesden:2010yp}. Moreover, SOR configurations are useful in constraining BBH formation channels~\cite{Gerosa:2013laa}. Recent studies have extensively investigated distinguishability  of the resonant spin configurations~\cite{Gupta:2013mea, Trifiro:2015zda, Gerosa:2014kta} as well as their detectability \cite{Afle:2018slw} through GW observations.

Recently, a  semi-analytical PN framework was developed to study the precession dynamics of spinning BBHs in quasi-circular orbits~\cite{Kesden:2014sla,Gerosa:2015tea,Gerosa:2016aus}. This framework utilizes the full timescale hierarchy of precessing binaries and constructs an effective potential, based on the mass-weighted effective spin parameter~\cite{Racine:2008qv}, to solve the 2PN orbit-averaged spin-precession equations analytically on $\tpre$. The solutions of these spin-precession equations provide relative orientations of $\Sa,\,\Sb$ and $\mathbf{L}$ in terms of a single parameter, total spin magnitude $S = \mid \Sa + \Sb \mid$.
These {\it one-parameter} orientations are then used to construct precession-averaged radiation reaction equations that are much faster to evolve than the equations in the orbit-averaged approach. Numerical implementation of this framework is available in the open-source package PRECESSION~\cite{Gerosa:2016sys}. In this effective potential based framework, the spin precession dynamics is classified into three mutually exclusive morphologies that encode the phenomenology of spin precession. The probability of a BBH being in one of these spin morphologies at a particular orbital separation depends on the orientations of BH spins during their formation; thus spin morphologies are indicative of binary's formation history \cite{Gerosa:2018wbw,Gerosa:2015tea}. The measurements of morphologies of a population of BBHs using GWs can provide valuable physical and astrophysical insights into their formation~\cite{Gerosa:2018wbw}. 
The three spin morphologies are categorized by the characteristic evolution of the difference in azimuthal angles of $\Sa$ and $\Sb$ on to the orbital plane, namely $\Delta\Phi$, in a precession cycle and comprise of two resonant morphologies where $\Delta\Phi$ librates around $0^\circ$ and $180^\circ$, and one circulating morphology where $\Delta\Phi$ sweeps through $0^\circ-180^\circ$ on $t_{pre}$ \cite{Kesden:2014sla}. The SOR configurations of Ref.~\cite{Schnittman:2004vq} are the extreme configurations in the two resonant morphologies in which the oscillation of $\Delta\Phi$ vanishes at $0^\circ$ and $180^\circ$. 

To date, studies of spin precession dynamics have mostly focused on BBHs in circular orbits. This might have been the case because spin precession effects are dominant only during the late stages of inspiral, and by that time BBHs formed with non-zero eccentricity are circularized due to the loss of energy and angular momentum in the form of GWs~\cite{Peters:1963ux,Peters:1964zz}. Notwithstanding this canonical wisdom, it was shown in Refs.~\cite{Klein:2010ti,Jain:2019ajn} that owing to 2PN spin-spin interactions, the orbital eccentricity can grow in the late stages of inspiral after reaching a minimum. Moreover, there exists disparity, particularly on eccentricity evolution, among different methods for solving the two-body problem in PN formalism (see Ref.~\cite{Loutrel:2018ydu,Will:2019lfe} for review). The recent discovery of strong secular growth in eccentricity obtained by solving  two-body PN equations using the osculating method~\cite{Loutrel:2018ssg} contrasts with the monotonic decay in eccentricity obtained using the orbit-averaged approach to PN approximation.  Eccentricity growth in extreme mass ratio binaries has also been seen within the self-force formalism~\cite{Cutler:1994pb}.  
Furthermore, many population synthesis studies show formation of considerable fraction of compact binaries with high eccentricity whose GWs would be in the frequency band of aLIGO-type detectors with non-negligible eccentricity~\cite{OLeary:2008myb,Samsing:2013kua,Samsing:2017jnz,Samsing:2017xmd,Samsing:2017plz,Samsing:2019ohp,Wen:2002km}. In such a scenario, it is worth investigating the spin dynamics of compact binaries in eccentric orbits.

In this paper, we use the effective potential based formalism of Ref.~\cite{Kesden:2014sla} to study the precession dynamics of spinning BBHs in eccentric orbits. We apply the spin morphology classification on binaries in eccentric orbits and evolve them from an initial separation ($a=1000\mathbf{M}$) to near merger ($a=10\mathbf{M}$) using PN accurate equations for spins, orbital angular momentum, and eccentricity. We observe three distinctively different evolution patterns for eccentricity in BBHs which mainly depend on their initial eccentricities and spins, $\Sa$ and $\Sb$: (i) eccentricity monotonically increasing until final orbital separation, (ii) eccentricity rising after decaying to a minimum value, and (iii) eccentricity monotonically decreasing throughout the inspiral. 
Since spin magnitudes affect the precessional dynamics as well as the eccentricity evolution, we also study the  morphology transitions of precessing BBHs in eccentric orbits. We track the morphology of the above three populations of BBHs with distinct eccentricity evolution and find that statistically the number of BBHs transiting from one morphology to other does not get affected by the presence of eccentricity in the binary dynamics. This finding, i.e., the statistical independence of morphology transition from eccentricity, is remarkable as it suggests that the morphology classification of precessing BBHs, initially developed for binaries in quasi-circular orbits, can also be used to probe the formation channels of the binaries with arbitrary eccentricity.

The rest of the paper is organized as follows. In Sec.~\ref{subsec:morph}  we describe the effective potential based PN framework to study precession dynamics of generic spinning BBHs. This formalism uses the evolutionary pattern of $\Delta\Phi$ in a precessional period to classify the dynamics of BBHs into three mutually exclusive spin  morphologies. 
In Sec.~\ref{sec:pn-eg},  we review the PN evolution equations for spins and the orbital elements used to evolve BBHs from a large orbital separation to near merger. In Sec.~\ref{sec:ecc_evol} we study the evolution of eccentricity in precessing BBHs during the inspiral. 
At each instantaneous separation, we employ the effective potential based framework to classify the spin dynamics of BBHs in eccentric orbits. Section \ref{results} shows our results for the morphology transitions in eccentric precessing BBHs. We conclude the paper in Sec.~\ref{sec:conclusion}.

\section{Methods}\label{sec:method}
\subsection{Notation} 
Precessing BBHs are, in general, characterized by the following physical parameters: the mass ratio $q = m_2/m_1 \le 1$, where $m_i$ ($i=1,2$) denote the component masses, the six components of their two spin angular momenta $\Si$, where spin magnitudes $\mid\Si\mid = m_i^2 \chi_i$ are parameterised  by dimensionless spin magnitudes  $0 \leq \chi_i \leq 1$, and eccentricity $e$. We consider  the total mass of BBHs,  $M =m_1+m_2 =1$, as it sets the overall scale in general relativity. The symmetric mass ratio of the system under consideration is denoted by $\eta=m_1m_2/M$. The total spin of a BBH is given as $\vek{S}=\vek{S}_1+\vek{S}_2$. The mean motion $n$ of eccentric BBH is related to the semi-major axis $a$ and orbital period (pericenter to pericenter) $P$  at the leading order by  the relation, $n = 2\pi/P = a^{-3/2}$~\cite{Klein:2010ti}.  In terms of the mean motion $n$, the PN expansion parameter can be expressed as $x = n^{2/3}$.  Throughout the paper, we will work in geometric units ($G\,=\,c\,=\,1$). 

\subsection{Morphological classification of precessional dynamics}\label{subsec:morph}
In this section, we briefly review the PN framework developed in Refs.~\cite{Kesden:2014sla,Gerosa:2015tea,Gerosa:2016aus} to study precession dynamics of spinning BBHs in quasi-circular orbits. This framework is meant for computing analytical solutions to 2PN orbit-averaged spin-precession equations on  $t_{pre}$ and then construct a set of precession-averaged evolution equations for BBHs inspiralling in circular orbits.  Henceforth, we shall refer to this formalism as circular orbit (CO) formalism and the binaries in circular orbits as CO binaries.  This framework exploits conservation of numerous physical quantities to construct the parametrized solutions of the orientation of spins $\vek{S}_i$ and the orbital angular momentum $\vek{L}$ on precessional time $t_{pre}$. In precessing binaries, the three angular momenta $\vek{L}, \, \vek{S}_1,\, \vek{S}_2$ precess around the total orbital angular momentum $\vek{J}$, constituting a nine-dimensional parameter space. The CO framework utilizes the conservation of $\vek{J}$ and the magnitude of $\vek L$ on $\tpre$, conservation of spin magnitudes $S_i$ on both $\tpre$ and $\trr$~\cite{ Apostolatos:1994mx, Buonanno:2005xu}, and the conservation of projected effective spin $\xi = M^{-2}\,[(1+q)\Sa + (1+q^{-1})\Sb]\cdot\hat{\mathbf{L}}$ by  both the 2PN orbit-averaged spin-precession equations and 2.5PN radiation reaction equations~\cite{Damour:2001tu, Racine:2008qv}. These conserved quantities reduce the degrees of freedom of precession motion from nine to two. In a suitable frame of reference, the relative orientations of $\Si$ and $\vek L$ can be parameterized by a single parameter, namely the total spin magnitude $S$.  Conservation of the projected effective spin $\xi$ on $\tpre$ is the nucleus of the formalism, which motivated the construction of two effective potentials $\xi_\pm(S)$ in the parameter space of spins. The effective potentials are defined as,
\begin{equation} \label{E:EP}
\xi_\pm(S)=\{ (J^2-L^2-S^2)[S^2(1+q)^2-(S_1^2-S_2^2)(1-q^2)]\pm(1-q^2)A_1A_2A_3A_4\} /(4qM^2S^2L).
\end{equation}
where,
\begin{subequations} \label{E:A}
      \begin{eqnarray} 
              A_1 &=& [J^2-(L-S)^2]^{1/2}\, , \\ 
              A_2 &=& [(L+S)^2-J^2]^{1/2} \, , \\ 
              A_3 &=& [S^2-(S_1-S_2)^2]^{1/2}\, , \\ 
              A_4 &=& [(S_1+S_2)^2-S^2]^{1/2}.
              \end{eqnarray}
\end{subequations}
For generic unequal mass BBHs, the two effective potentials $\xi_\pm(S)$ form a loop in $ S-\xi$ space (e.g., see Fig.~1 in Ref.~\cite{Kesden:2014sla}). In a precession period, the total spin magnitude $S$ oscillates along a horizontal line between two turning points $S_+$ and $S_-$ which lie on the $\xi_+(S)$ and $\xi_-(S)$ curve, respectively. While the preceding statement is true for freely precessing BBHs, for binaries near the SOR configurations both the turning points can lie on either $\xi_+(S)$ or $\xi_-(S)$ curve.
At the extrema of the loop, the turning points are degenerate, and these two points in the $S-\xi$ loop correspond to the two SOR configurations in BBHs. 
Since the orientations of $\Si$ are parameterized by the single parameter $S$, the spin-precession dynamics on $\tpre$ can be studied by simply evolving the angular parameters of $\Si$ between $S_+$ and  $S_-$. The three angles $\theta_i = \arccos (\hat{\Si}\cdot\hat{\vek{L}})$ and $\theta_{12}=\arccos(\hat{\Sa}\cdot\hat{\Sb})$ evolve monotonically over half the precession cycle. The angle $\Delta\Phi=\arccos \left[\left( \hat{\Sa} \times \hat{\vek{L}}\Big /|\hat{\Sa} \times \hat{\vek{L}}| \right)\cdot \left(\hat{\Sb} \times \hat{\vek{L}}\Big /|\hat{\Sb} \times \hat{\vek{L}}\right)\right]$, on the other hand, evolves characteristically depending on the values of $\Si,\, \vek{L}$, and $\vek{J}$ of the binary~\cite{Gerosa:2015tea}. Three qualitatively different evolution of $\Delta\Phi$ on $\tpre$ provide a unique geometric way to classify the spin precession dynamics in the following three, mutually exclusive, morphologies:

\begin{enumerate}[label=\Roman*.]
\item Circulating morphology (\textbf{C}): Circulation of $\Delta\Phi$ between [$-\pi, \pi$].
\item Librating morphology about $0$ ($\vek{L0}$): Oscillation of $\Delta\Phi$ about 0.
\item Librating morphology about $\pi$ ($\vek{L\pi}$): Oscillation of $\Delta\Phi$ about $\pi$.
\end{enumerate}

The BBHs in the {\textbf C}-morphology correspond to the freely precessing binaries, whereas binaries in $\vek{L0}$ and  $\vek{L\pi}$ morphologies are librating about the planar configurations of $\Si$ and $\vek{L}$.  The two types of  SOR configurations \cite{Schnittman:2004vq}: $0$-SOR ($\Si$ and $\vek{L}$ being in a plane with $\Delta\Phi=0$) and $\pi$-SOR ($\Si$ and $\vek{L}$ being in a plane with $\Delta\Phi=\pi$) fall in the $\vek{L0}$-morphology and $\vek{L\pi}$-morphology, respectively. The SOR configurations are important for an understanding of precession dynamics. Previous studies have shown that the precessional dynamics can be explained in terms of proximity of the spin configurations to the SOR configurations~\cite{Gupta:2013mea, Schnittman:2004vq}. These studies have shown that inspiralling BBHs near the SOR configurations eventually get captured in the SOR configurations or oscillate about the SOR configurations during the course of gravitational radiation driven evolution thereby leaving a characteristic imprint on the distributions of final spins.  The spin configurations of BBHs at a particular orbital separation represent a snapshot of BBHs that are undergoing precession on $\tpre$. The identification of spin morphologies complements these studies, which describe the average behavior of BBHs' spins on a precessional cycle. The morphologies remain constant on $\tpre$ and slowly evolve under radiation reaction. The uncertain number of precession cycles between BBH formation and merger implies that the spin angles $\theta_1$, $\theta_2$ and $\Delta \Phi$ near merger
cannot be predicted from the initial conditions in practice. However,
the precession-averaged equation provided in Refs. \cite{Kesden:2014sla,Gerosa:2015tea} can be used to predict the
spin morphology near merger with confidence since it is evolving on
the slower radiation-reaction time $t_{rr}$.


\begin{figure}[!ht]
    \centering
    \includegraphics{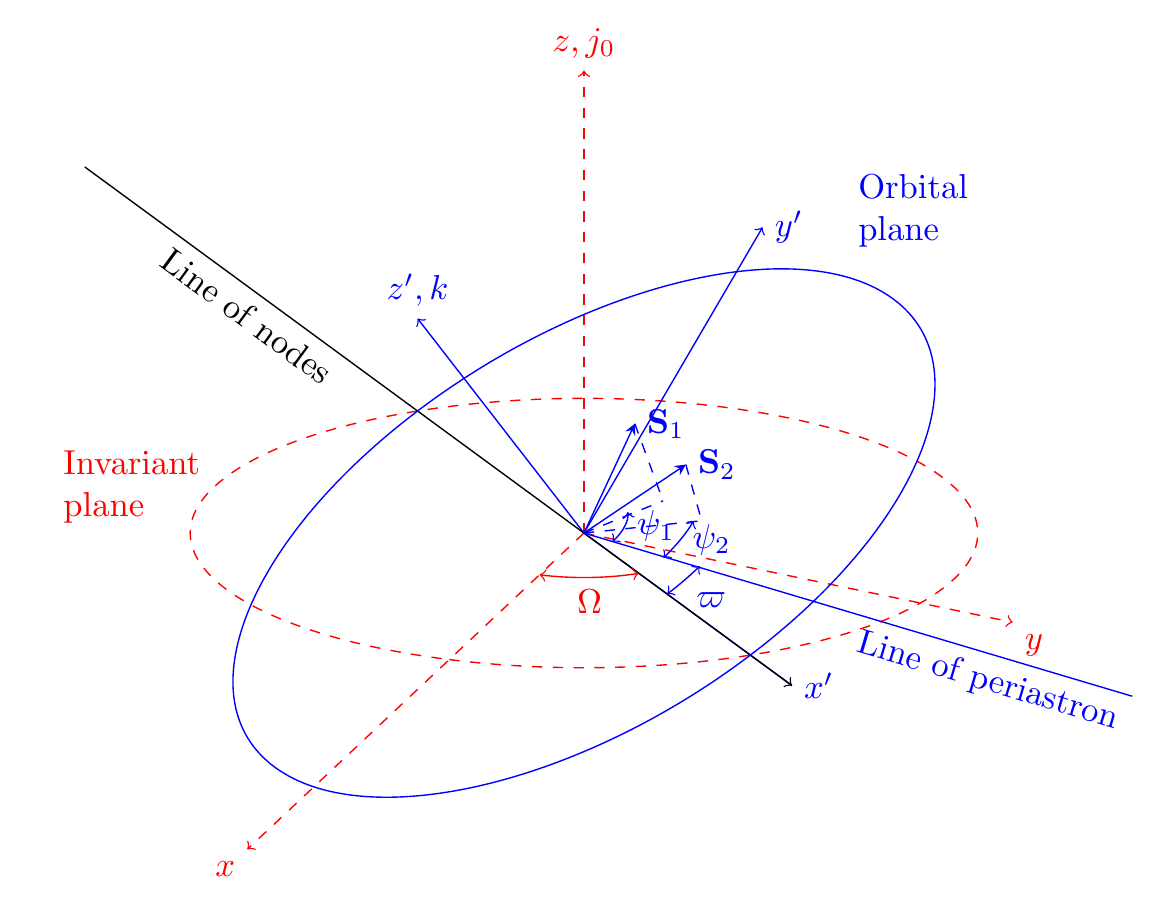}
    \caption{\label{fig:ref-frame}Reference frames used and various angles appearing in this study are shown in this figure. The inertial frame ($x,y,z$) is chosen such that the z-axis is along the direction of the total orbital angular momentum vector at $a\, = \,1000\mathbf{M}$ ($\mathbf{j_0}$ is the unit vector along the initial total orbital angular momentum at $a=1000\mathbf{M}$). The invariant plane is on the x-y plane of the inertial frame. The orbital plane of binary is spanned by $x^\prime-y^\prime$ plane of the co-precessing frame ($x^\prime, y^\prime, z^\prime$) whose $x^\prime$-axis and $z^\prime$-axis are along the line of nodes and the orbital angular momentum vector ($\mathbf{k}$ is the unit vector along the orbital angular momentum), respectively. The angles $\psi_1$ and $\psi_2$ are the angles between the line of periastron and projection of $\Sa$ and $\Sb$ on the orbital plane, respectively. The longitude of $x^\prime$-axis or line of nodes in the invariant plane is given by $\Omega$.}
\end{figure}

\subsection{Post-Newtonian evolution equations }\label{sec:pn-eg}

In generic binaries, if the spins are not aligned or anti-aligned with the orbital angular momentum, the spins and the orbital plane precess about the total angular momentum.  At large separations, the spin-induced changes in orientations of spins and orbital angular momenta are much slower than the orbital periods. Using this fact, the evolution of angular momenta vectors can be described by averaging over the instantaneous changes  occurring in orbital time $t_{orb}$. The 2PN orbit-averaged equations describing the evolution of spins and orbital angular momentum vectors are given as~\cite{Barker:1975ae,Racine:2008qv},

\begin{subequations} \label{E:SP}
      \begin{eqnarray} \label{E:SP1}
              \frac{d\Sa}{dt} &=&  \frac{1}{2a^3(1-e^2)^{3/2}}\left[ \left( 4 +  3q- \frac{3\,( \Sb+q\,\Sa )\cdot \vek{L}}{L^2}\right) \vek{L}\, + \Sb  \right] \times \Sa \, , \\ \label{E:SP2}
                  \frac{d\Sb}{dt} &=& \frac{1}{2a^3(1-e^2)^{3/2}}\left[ \left( 4 + 3q^{-1} -\frac{3\,( \Sa + q^{-1}\,\Sb )\cdot\vek{L}}{L^2} \right)\vek{L}\,+\Sa \right] \times \Sb \, ,\\
                  \frac{d\vek{L}}{dt}&=&  \pmb{\omega}_p \times \vek{L}\,,
                    \end{eqnarray}
\end{subequations}

where $\vek{L}$ is the Newtonian orbital angular momentum vector while $\pmb{\omega}_p$ is given as,

\begin{subequations}\label{E:preces-freq}
    \begin{eqnarray}
    \pmb{\omega}_p &=& \delta_1  \Sa + \delta_2 \Sb\label{e:prec-vec}\,,\\
    \delta_1 &=& \frac{1}{2a^3(1-e^2)^{3/2}} \left( 4 + 3\,q - \frac{3\,( \Sb+q\,\Sa )\cdot \vek{L}}{L^2} \right) \,,\\
    \delta_2 &=& \frac{1}{2a^3(1-e^2)^{3/2}} \left( 4 + 3\,q^{-1} - \frac{3\,( \Sa+q^{-1}\,\Sb )\cdot \vek{L}}{L^2} \right) \,.
    \end{eqnarray}
\end{subequations}

In the absence of gravitational radiation, the magnitude of $|\vek{L}|$ which depends upon the orbital elements $e$ and $a$ remains a constant while its orientation changes in the precession timescale. The spin and orbital angular momentum vectors evolve in a much longer time than the orbital time in the PN regime, allowing to describe the orbital motion using quasi-Keplerian parametrization on the orbital plane~\cite{SCHAFER1993196, Damour:1988mr}. The quasi-Keplerian formalism provides analytical solutions to the conservative part of PN equation of motion of binaries as functions of the eccentric anomaly ``$u$". The quasi-Keplerian parametrization has been derived to various PN orders for non-spinning as well as spinning binaries in elliptical orbit~\cite{Konigsdorffer:2006zt, Memmesheimer:2004cv, Klein:2010ti}. A spinning binary's orbit at 2PN order quasi-Keplerian description can be expressed as,
\begin{subequations} \label{E:qk1}
\begin{eqnarray}
r &=& a(1-e_r \cos u) + f_r(v)\,,\\
\phi &=& (1+k)v + f_\phi(v)\,, \\
v &=& 2\arctan\left(\sqrt{\frac{1+e_\phi}{1-e_\phi}} \tan \frac{u}{2}\right)\,,\\
l &=& u - e_t \sin u + f_t \left( u, v\right) \,,\\
\dot{l} &=& n\,,
\end{eqnarray}
\end{subequations}
where $(r, \phi)$ are polar coordinates of separation vector in the orbital plane; $u, v$ and $l$ are the eccentric, true, and mean anomalies; and $k$ accounts for periastron precession. The functions $f_i$, appearing in above equations are 2PN accurate orbital functions~\cite{Klein:2018ybm, Klein:2010ti, Damour:2004bz, Memmesheimer:2004cv}. For brevity, we have not explicitly listed the expressions of the PN orbital correction functions as these are of little importance in this paper. Our expressions of the PN orbital corrections functions match with those in Ref.~\cite{Klein:2018ybm}, where quadrupole-monopole interaction terms are incorporated.  The quasi-Keplerian parametrization introduces three eccentricity parameters $e_t,~e_\phi$ and $e_r$, all the three eccentricities are related to each other, differ from each other in PN correction terms~\cite{Damour:2004bz, Memmesheimer:2004cv}. 
 The  parametrized solution of orbital phase  can be further decomposed into a linear part $\lambda$, often termed as mean orbital phase, and an oscillatory part $W_\phi$~\cite{Damour:2004bz} as

\begin{subequations}\label{E:qk2}
\begin{eqnarray}
\phi &=& \lambda + W_\phi\label{E:phi-linear} \,,\\
\dot{\lambda} &=& (1+k)n \,,\\
W_\phi &=& (1+k)(v-l) + f_\phi(v)\,.
\end{eqnarray}
\end{subequations}

Once radiation reaction is included, the binary evolves slowly in radiation-reaction time $\trr$ and $|\vek{L}|$ decays. Consequently, the mean motion $n$, which is related to orbital frequency and eccentricity evolve according to the following equations~\cite{Damour:2004bz,Arun:2009mc,Klein:2018ybm},

\begin{subequations} \label{E:RR1}
	\begin{eqnarray}
		\frac{dn}{dt} &= &\eta \,x^{11/2}\left(  \dot{n}^{}_{\text{N}} + \dot{n}^{}_{1\text{PN}}\,x + \dot{n}^{}_{1.5\text{PN}}\, x^{3/2}+\dot{n}^{}_{2\text{PN}}\,x^2 +\dot{n}^{}_{2.5\text{PN}}\,x^{5/2}+\dot{n}^{}_{3\text{PN}}\,x^3\right)\,,\\
		\frac{de^2}{dt} &=& - \eta\, x^{4}\left(  \dot{e^2}^{}_{\text{N}} + \dot{e^2}^{}_{1\text{PN}}\, x + \dot{e^2}^{}_{1.5\text{PN}}\, x^{3/2}+\dot{e^2}^{}_{2\text{PN}}\, x^2 +\dot{e^2}^{}_{2.5\text{PN}}\, x^{5/2}+\dot{e^2}^{}_{3\text{PN}}\, x^3\right)\,.\label{eq:e2}
	\end{eqnarray}
\end{subequations}

The explicit expressions for various terms in the above equations are provided in Appendix \ref{appen:pn_coefficients}. Customarily, the eccentricity parameter in Eq.~(\ref{eq:e2}) is time eccentricity: $e \equiv e_t$. The evolution equations of mean motion $n$ and eccentricity $e$ have dependencies on the angles $\psi_1$ and $\psi_2$, which are subtended by the projections of spins $\Sa$ and $\Sb$ on the orbital plane from the line of periastron. These angles, shown in Fig.~\ref{fig:ref-frame}, bring secular effects of periastron advance in the evolution of binaries. The reference frame in which the angles are defined is co-precessing with the orbital plane of the binary, whose $x$-axis (hereafter $x^\prime$-axis) is in the direction of the line of nodes. The longitude of $x^\prime$-axis $(\Omega)$ evolves as binary precesses on precession time and changes at the same rate with the precession frequency of the orbital angular momentum vector~\cite{Konigsdorffer:2005sc}, given as

\begin{equation} \label{E:omega}
    \frac{d\Omega}{dt}= \omega_p,
\end{equation}
where $\omega_p=|\pmb{\omega}_p|$. The  two angles, mean orbital phase $\lambda$ and mean anomaly $n$ appearing in Eqs.~(\ref{E:qk2}) are defined relative to $x^\prime$-axis and the line of periastron, respectively. The difference between the two angles gives a measure of longitude of periastron line: $\varpi\,=\,\lambda - l$~\cite{Damour:2004bz, Moore:2016qxz, Klein:2018ybm}. As the binary evolves, $\lambda$ and $l$ drift apart in periastron precession timescale because of periastron advance. The evolution of $\varpi$ can be expressed as~\cite{Damour:2004bz, Klein:2018ybm},

\begin{equation}\label{E:varpi}
    \frac{d\varpi}{dt} = k\,n\,.
\end{equation}
The expression of $k$, that embodies the secular effect of periastron precession per orbital period can be written at the leading order as $k=3n^{2/3}/(1-e^2)$.  We recast the PN equations governing inspiral of generic spinning BBHs in eccentric orbits [Eqs.~(\ref{E:SP}) and (\ref{E:RR1})], evolution of $x^\prime$ [Eq.~(\ref{E:omega})], and evolution of $\varpi$  [Eq.~(\ref{E:varpi})]  in the following twelve equations

\begin{subequations}\label{E:RECAST}
	 \begin{eqnarray} 
	 \frac{de^2}{dn} &=&\frac{de^2}{dt}\Big/\frac{dn}{dt}\,,\\
	 \frac{d\Sa}{dn} &=&  \frac{d\Sa}{dt}\Big/\frac{dn}{dt}\,,\\
	 \frac{d\Sb}{dn} &=&  \frac{d\Sb}{dt}\Big/\frac{dn}{dt}\,,\\
	 \frac{d\vek{L}}{dn}&=& \frac{d\vek{L}}{dt}\Big/\frac{dn}{dt}\,,\\
	 \frac{d\Omega}{dn}&=&  \frac{d\Omega}{dt}\Big/\frac{dn}{dt}\,,\\
	 \frac{d\varpi}{dn}&=& \frac{d\varpi}{dt}\Big/\frac{dn}{dt}
	 \end{eqnarray} 
\end{subequations}

We simultaneously solve the above 12 ordinary differential equations using the explicit embedded Runge-Kutta Prince-Dormand (8,9) time integration scheme  with relative tolerance $10^{-8}$~\cite{DORMAND198019}. The initial configurations of  the generic spinning BBHs are generated at a separation $a = 1000\mathbf{M}$ which corresponds to the PN expansion parameter $x$ to be $10^{-3}$, and the directions of spin vectors are uniformly distributed over a sphere. For simplicity, we choose the argument of the line of periastron $\varpi$  at $a=1000\mathbf{M}$ to be zero, while the initial angle of the line of nodes or $x^\prime$-axis is set to be $45^\circ$.
We evolve the Cartesian components of $\Si$ and $\mathbf{L}$ in the inertial frame ($x,y,z$) all the way down to $a=10\mathbf{M}$, beyond which PN approximations become increasingly uncertain. The coupled ordinary differential equations depend on the angles $\psi_1$ and $\psi_2$, appearing in Eqs.~(\ref{eq:beta:sigma:tau}), which are defined in the orbital plane relative to the line of periastron.  To compute the angles, we used a dynamical mapping between the inertial frame ($x,y,z$) and co-precessing frame ($x^\prime, y^\prime, z^\prime$) using the Euler's angles: $\Omega$ and $\vartheta = \hat{\vek {L}} \cdot \hat{j_0}$ to get the azimuthal angles $\phi_i$ of individual spins in the co-precessing frame. Further, the azimuthal angles $\phi_i$ are subtracted by $\varpi$ at each orbital separation to get $\psi_i$ for respective spins.

While integrating the set of coupled ordinary differential equations, we exploited the fact that general relativity is scale free and set the total mass $M$ to unity.  Therefore, in our simulations, mass ratio $(q)$ is the only mass related intrinsic parameter. Each BBH at initial separation $a = 1000\mathbf{M}$ is specified by the mass ratio $q$, dimensionless spin magnitudes $\chi_i$, eccentricity $e$, and spherical coordinates of BH spins, $\left(\theta_1^\prime, \phi_1^\prime, \theta_2^\prime, \phi_2^\prime\right)$, in an inertial frame where components of the total orbital angular momentum $( \hat{J_x}, \hat{J_y},\hat{J_z})$  are $(0,0,1)$.
Since the precession dynamics is preserved under the rotation  of the spin component around the orbital angular momentum $\mathbf{L}$ in the orbital plane, we can describe the spin configurations of BBHs at any separation, without loss of generality, using only three angular coordinates $(\theta_1, \theta_2, \Delta\Phi )$ in the co-precessing frame where $\Delta\Phi=\phi_1-\phi_2$, $\theta_i$ and $\phi_i\, (i=1,2)$ are polar and azimuthal angles of respective spins, respectively.

\section{Evolution of orbital eccentricity}\label{sec:ecc_evol}

A number of definitions for eccentricity exist in the literature, resulting in different studies on the evolution of eccentricity on radiation reaction time-scale.  For example, Refs.~\cite{Konigsdorffer:2006zt, Klein:2010ti, Damour:2004bz} have used a variety of eccentricities to delineate generic orbits at various PN orders. In the osculating orbit formalism~\cite{Lincoln:1990ji}, the eccentricity and semi-major axis are defined in such a way that Keplerian orbit is momentarily tangent to the actual orbit. 
This osculating eccentricity is then expressed in terms of components of Runge-Lenz vector where a secular growth of eccentricity for non-spinning BBHs is observed~\cite{ Loutrel:2018ssg}. In fact, in numerical relativity simulations, several definitions of eccentricity and their extraction methods exist~\cite{Mroue:2010re,Berti:2006bj}. Numerical relativists also use eccentricity removal methods to construct quasi-circular initial data, which can reduce the eccentricity values to less than $10^{-4}$~\cite{Ramos-Buades:2018azo,Buonanno:2010yk}. A nice summary of these eccentricity definitions can be in Ref.~\cite{Loutrel:2018ydu}. In this paper, we adopt the quasi-Keplerian formalism of defining eccentricity as discussed in Refs.~\cite{Konigsdorffer:2006zt, Klein:2010ti, Damour:2004bz} and 
limit ourselves to studying the evolution of only `temporal' component of the eccentricity.
Heretofore, by ``eccentricity" we will always mean this component; hence, $e  \equiv e_t$.

We evolve generic spinning BBHs in eccentric orbits from an initial separation ($1000\mathbf{M}$) to late inspiral ($10\mathbf{M}$) for the following three scenarios: 
\begin{enumerate}[label=\Roman{*}., ref=(\roman{*})]
\item eccentricity  monotonically increases from the aforementioned initial orbital separation through the final orbital separation,
\item eccentricity rises after initially decaying to a minimum value,
\item eccentricity monotonically decreases throughout the inspiral.
\end{enumerate}

\begin{figure}[!ht]
\centering
\includegraphics[width=.8\textwidth]{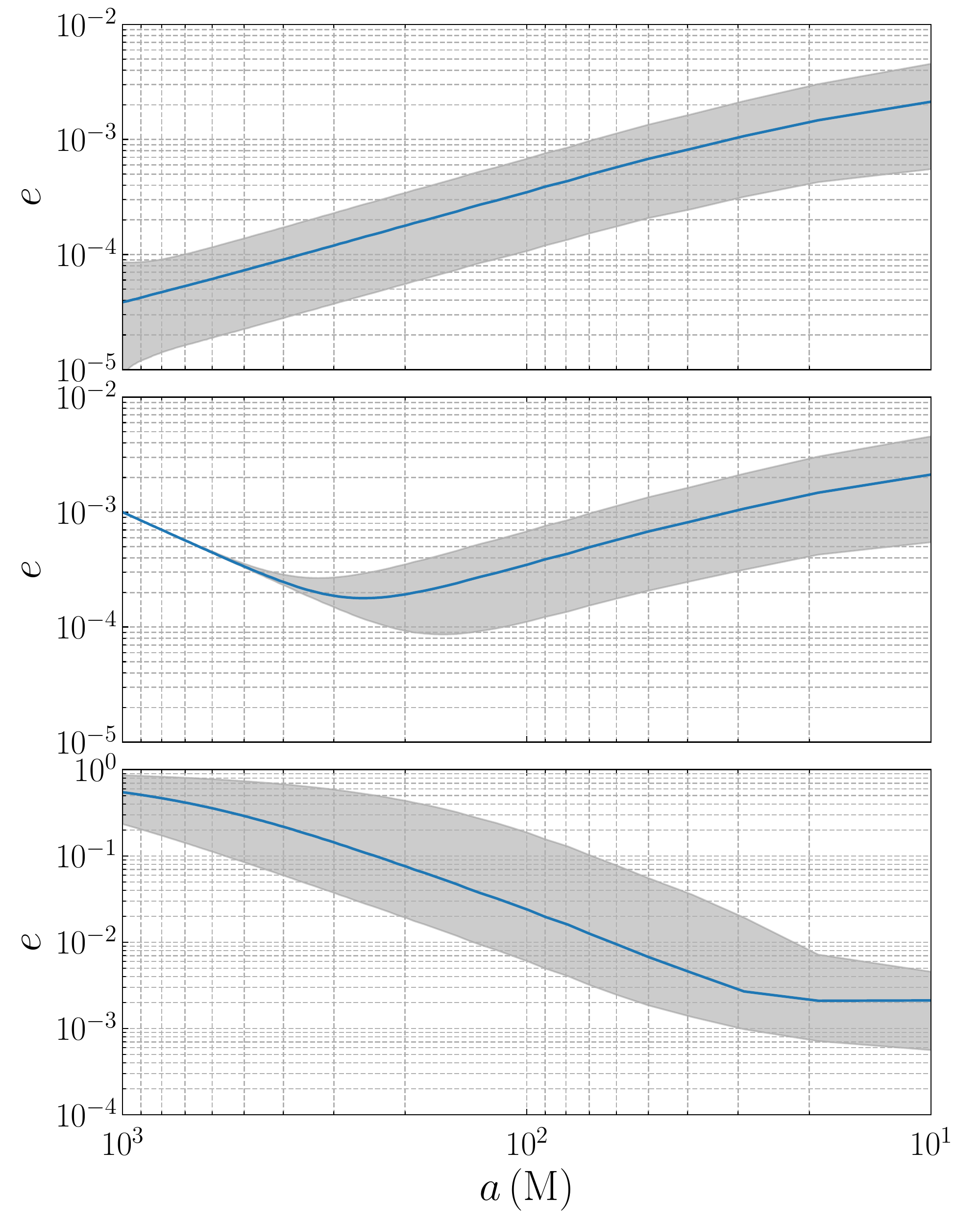}
\caption{\label{Fig:ecc_evol}The figure depicts three different patterns of eccentricity evolution from an initial separation $a\,=\,1000\mathbf{M}$ to a final separation $a\,=\,10\mathbf{M}$ for a population of BBHs.  In each panel, the grey region is between the 5th and the 95th percentile of eccentricities.
The solid lines in the three panels represent the median of eccentricities. In the top panel, the eccentricities monotonically increase throughout the inspiral. The initial eccentricities  in the first panel correspond to the values where the derivative $de^2/dt$ vanishes at 2PN order. In the middle panel, eccentricities rise after decaying to a minimal value. In the bottom panel, we notice the canonical monotonic decay of eccentricities.}
\end{figure}

The three different types of eccentricity evolution are shown in Fig.~\ref{Fig:ecc_evol}.  To study the spin dynamics of BBHs in precession time $\tpre$ at any arbitrary separation, we construct the angular parameters of spins $(\theta_1, \theta_2, \Delta\Phi)$ from the evolved Cartesian components of $(\Sa, \Sb, \mathbf{L})$ at the instantaneous frame $(x^\prime, y^\prime, z^\prime)$ where $( \hat{L_x}=0, \,\hat{L_y}=0,\,\hat{L_z}=1)$  and then employ the morphology-based classification scheme of spin dynamics, discussed in detail in the next section. 
Note that the basis of this scheme of classifying spin dynamics in different morphologies is conservation of $\xi$ in precession period. The effective spin $\xi$ is also preserved for eccentric binaries by virtue of spin-precession equations (Eqs.~(\ref{E:SP})) and marginally preserved during inspiral\footnote{We check that the median change in the value of $\xi$ from $1000\mathbf{M}$ to $10\mathbf{M}$ for binaries considered in this paper is ${\cal O}(10^{-4})$.}, implying that the morphology classification formalism can be trivially extendable to eccentric binaries. For comparison purposes, we also evolve spinning BBHs in circular orbits and compute their morphologies using the python package called PRECESSION~\cite{Gerosa:2016sys}.

In this section, we investigate the effect of orbital eccentricity on the precession dynamics of BBHs.  We ran three different sets of simulation based on the types of eccentricity evolution mentioned above. The spin precession induces nontrivial evolution of eccentricity in these sets of simulation.  In the first set, the eccentricities monotonically increase throughout the inspiral, as shown in the top panel of Fig.~\ref{Fig:ecc_evol}. The initial eccentricity of these BBHs at separation $a=1000\mathbf{M}$ correspond to the values $e_{\rm min}$ where the projections of two spins in the orbital plane cancel the derivative $de^2/dt$  at 2PN order~\cite{Klein:2018ybm,Klein:2010ti}. The minimum eccentricity $e_{\rm min}$ depend on the spin orientations. The 2PN spin effect stops further decay of eccentricity beyond $e_{\rm min}$. In this set of simulations, the spins vectors $\Sa$ and $\Sb$ are uniformly distributed over a 2-sphere and the dimensionless spin magnitudes are chosen to be $\chi_1,\chi_2 \in \{ 0.2,\,0.6\,,1.0\}$  with mass ratios $q\in\{0.4\,, 0.6\,, 0.8\,, 0.95 \}$. 

In the second set of simulations, the eccentricities  recuperate back after decaying to their respective minimum values $e_{\rm min}$, where the spin-spin coupling starts inducing positive slope in $de^2/dt$ (Eq.~\ref{eq:e2}). This particular evolutionary pattern of eccentricity is shown in the middle panel of  Fig.~\ref{Fig:ecc_evol}. In this simulation set, the initial eccentricities of all the BBHs are fixed to be $e_{\rm ini}=0.001$ while the mass and spin parameters at $a=1000\mathbf{M}$ are same as in the first set.
In the third set of simulations, the eccentricities of BBHs show the canonical decaying pattern. For this set, the initial eccentricities of the binaries at $a=1000\mathbf{M}$ are sampled from a uniform distribution between $e_{\rm ini}=0.2$ and $e_{\rm ini}=0.9$ while the other intrinsic parameters of the BBHs are distributed in the same manner as in the first and second sets.

In each set of simulations, we evolve \numprint{36000} BBHs over the parameter space $\left( q\,, e\,, \Sa, \Sb \right)$. The minimum eccentricity values $e_{\rm min}$ of BBHs, where eccentricity ceases to decay during inspiral, depends only on the spin-spin coupling or the spin magnitudes $\left(\chi_1, \chi_2\right)$ whereas they are almost independent of the mass ratio $q$. This is because 
for  given values of $\left( e_{\rm ini}, \chi_1, \chi_2  \right)$, the eccentricities follow roughly similar evolution for all mass ratios $q$ considered in this work.  In the top panel of Fig.~\ref{Fig:ecc_evol}, we show monotonic rise of eccentricities where the initial eccentricities of BBHs are $e_{\rm ini} = e_{\rm min}$. In the middle panel, another nontrivial pattern of eccentricity evolution is shown where eccentricities decay to $e_{\rm min}$ before spin-spin interaction induce a rise of eccentricities.  The late inspiral growth of eccentricity is observed in BBHs with initial eccentricity $e_{\rm ini} \le 0.1$.  The precise value of orbital separation $a$, where minimum eccentricity occurs depends on the choice of parameters ($e, \Sa, \Sb$). The precession induced growth of eccentricity is different than the growth of eccentricity in the late inspiral observed in Refs.~\cite{Loutrel:2018ssg,Cutler:1994pb}. In Fig.~\ref{Fig:ecc_evol}, the grey regions represent eccentricities of BBHs between the 5th and the 95th percentile of their populations. The width of the grey region is attributed to varying spin-spin coupling strengths or different spin magnitudes $\left(\chi_1, \chi_2\right)$ across the BBH population.

\begin{figure}[ht!]
        \includegraphics[width=\textwidth]{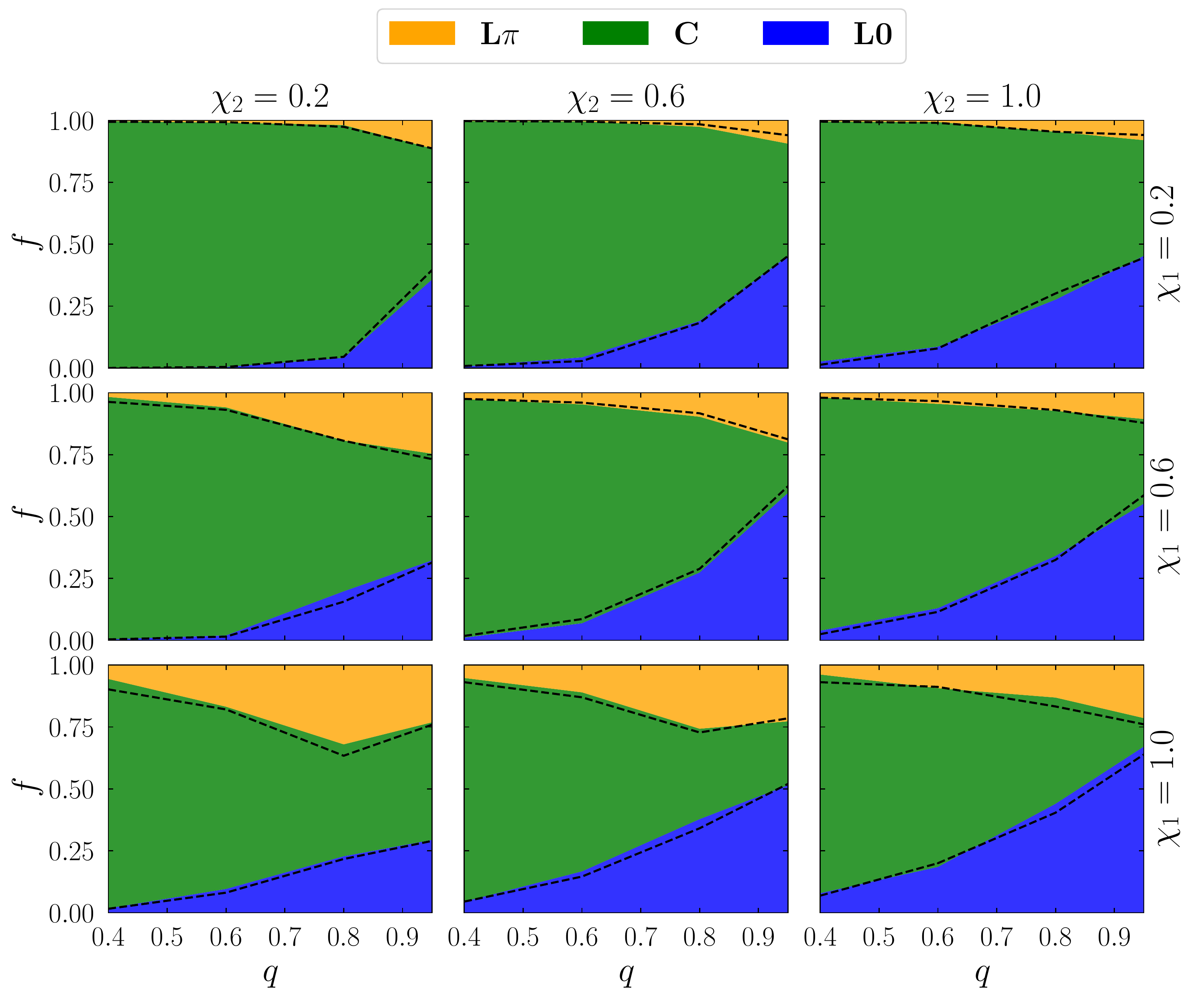}\\
    \caption{ \label{Fig:morph_mono_increase}Fraction $f$ of BBHs  in the three different spin morphologies at separation $a=10\mathbf{M}$ is shown as a function of  mass ratio $q$ and spin magnitudes $(\chi_1, \chi_2)$.  Here, eccentricities of BBHs are monotonically increasing during the inspiral (see the top panel of Fig.~\ref{Fig:ecc_evol}). 
    The yellow, green, and blue colored patches represent the fraction of BBHs in $\vek{L0}$-morphology, \textbf{C}-morphology, and $\vek{L\pi}$-morphology, respectively. For each combination of $(q, \chi_1, \chi_2)$, the spin orientations of BBHs are distributed isotropically at the initial separation $a=1000\mathbf{M}$. The black dashed black lines represent the boundaries of different morphologies for BBHs in circular orbits and have been plotted to compare the evolution of spin morphologies of BBHs in circular orbits with that of BBHs in eccentric orbits.  This plot shows that the presence of eccentricity has no significant impact on the transitions of BBHs to different morphologies during the inspiral.}
\end{figure}

\begin{figure}[!hb]

        \includegraphics[width=\textwidth]{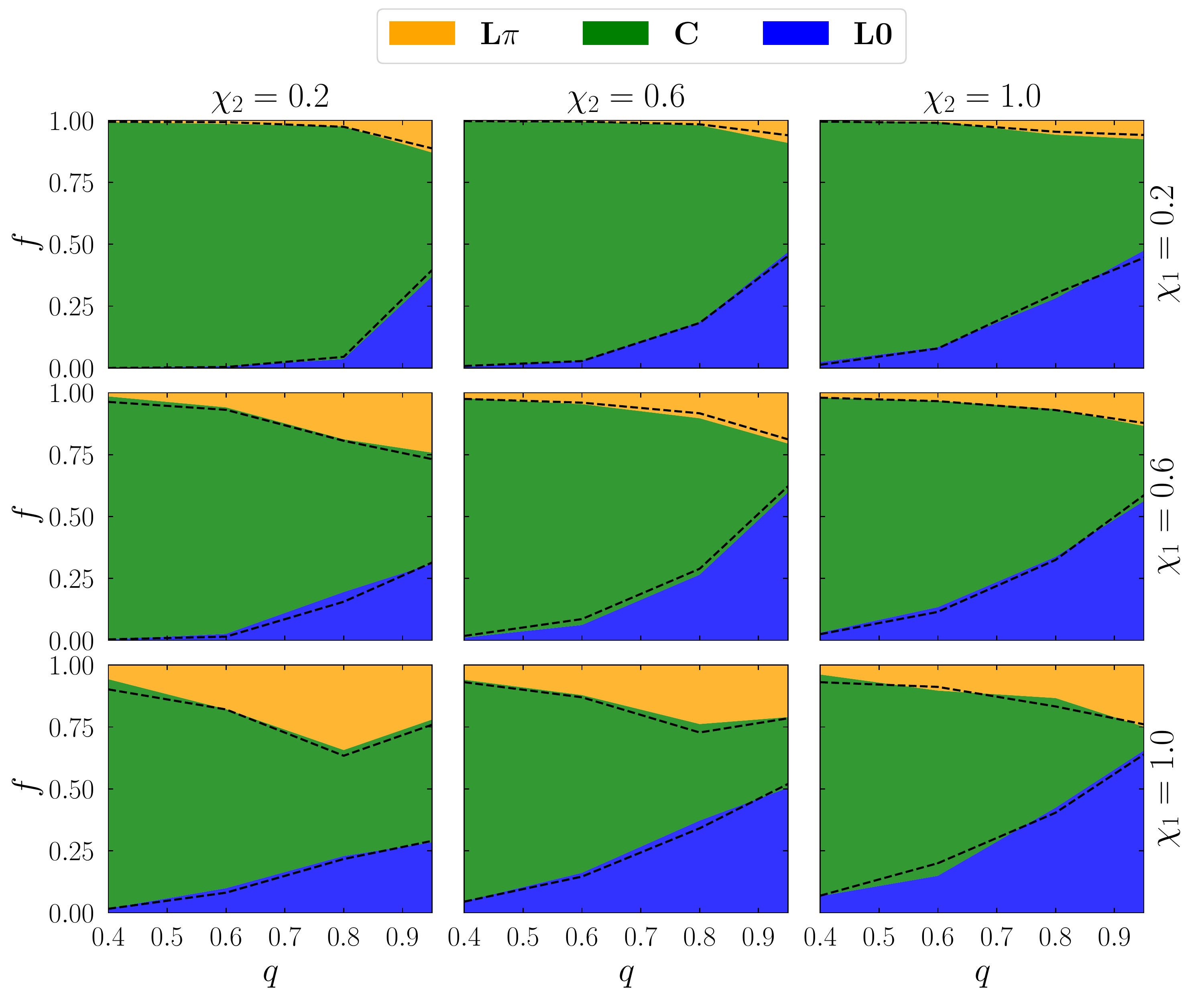}

    \caption{\label{Fig:morph_decrease_rise} Same as in Fig.~\ref{Fig:morph_mono_increase}, but for the case depicted in the second panel in Fig.~\ref{Fig:ecc_evol}, where the BBH eccentricities grow after decaying to certain minima $e_{\rm min}$. BBHs in this plot have their spins isotropically distributed at the initial separation $a=1000\mathbf{M}$.  The fraction $f$ of BBHs with eccentricity in different morphologies is not different from that of BBHs in circular orbits in those morphologies. The colored patches and the dashed lines have the same meaning as in Fig.~\ref{Fig:morph_mono_increase}.}
\end{figure}

\begin{figure}[ht!]
\includegraphics[width=\textwidth]{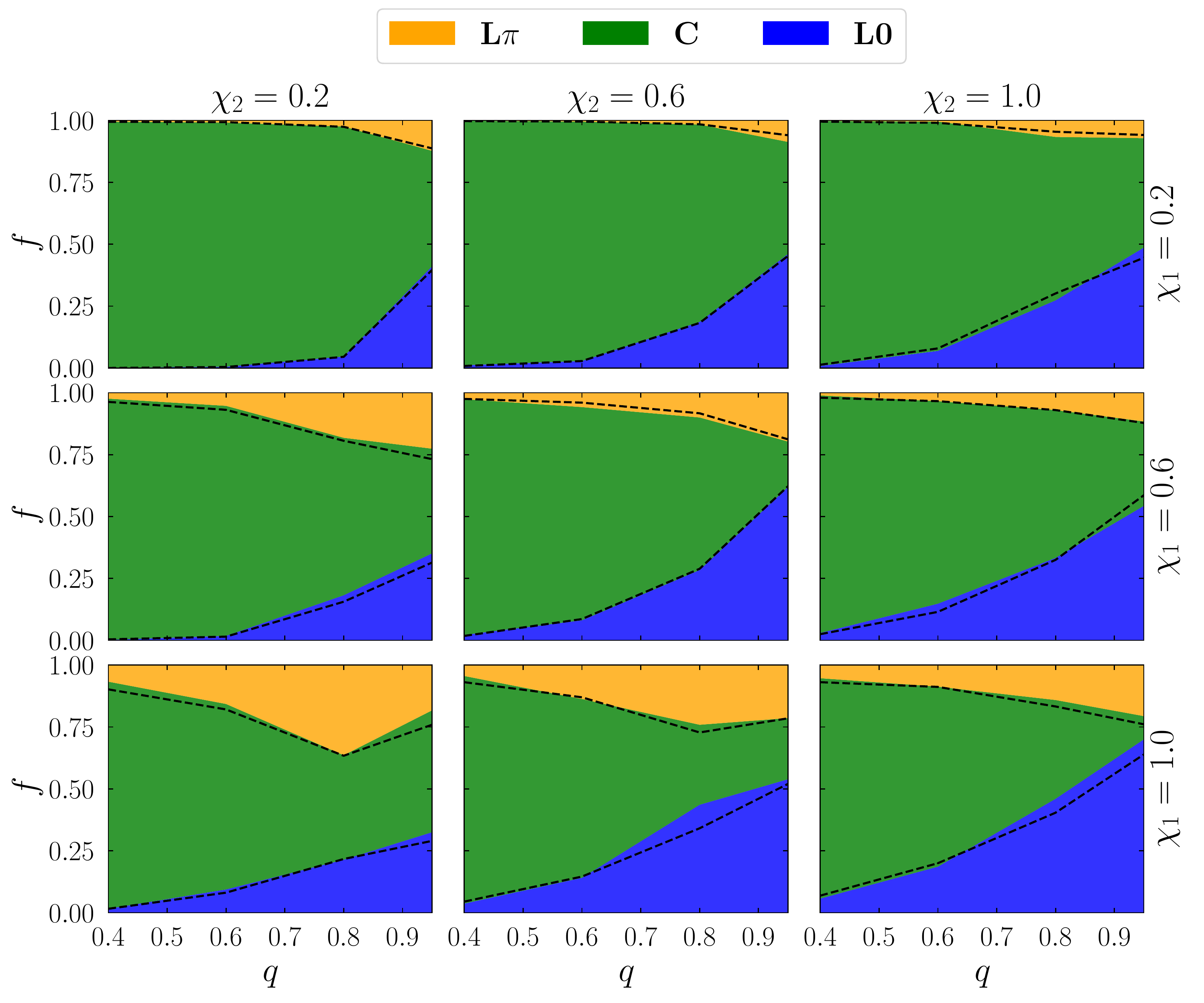}
\caption{\label{Fig:morph_mono_decay} Eccentricities of all BBHs in this figure exhibit canonical monotonic decay during inspiral, which is the case shown in the bottom panel in Fig.~\ref{Fig:ecc_evol}.  The colors have the same meaning as in Figs.~\ref{Fig:morph_mono_increase} and \ref{Fig:morph_decrease_rise}.  The dashed lines denote the  boundaries of spin morphologies of BBHs in circular orbits at $a=10\mathbf{M}$.  In the bottom-middle panel, the differences between the fraction of eccentric BBHs getting captured in the spin morphologies and the fraction of circular BBHs in respective morphologies are not within Poisson counting error bound at $\left( q = 0.8, \chi_1 = 1.0, \chi_2 = 0.6\right)$.}
\end{figure}

\section{Spin Morphology of Eccentric BBHs}
\label{results}

It has been shown that with the increase in BH spin magnitudes, the number of BBHs in circular orbits transiting from $\mathbf{C}$-morphology to resonant morphologies, $\mathbf {L0}$ and $\mathbf{L\pi}$~\cite{Gerosa:2015tea} increases. Since the eccentricity evolution is correlated with spin magnitudes \cite{Klein:2010ti}, we study the effects of eccentricity on the binary precession dynamics and their morphology transition. For a qualitative understanding, particularly given the distinctive late inspiral evolution of eccentricity, we compute the spin morphologies of the three sets of BBH populations having different eccentricity evolution patterns as shown in Fig.~\ref{Fig:ecc_evol}. In Figs.~\ref{Fig:morph_mono_increase}, \ref{Fig:morph_decrease_rise} and \ref{Fig:morph_mono_decay}, we plot the fraction of binaries in eccentric orbits  in different spin morphologies as a function of mass ratio $q$ and spin magnitudes $(\chi_1, \chi_2)$ at orbital separation $a=10\mathbf{M}$. Different color patches represent regions of three different morphologies: green for $\mathbf{C}$, blue for $\mathbf{L0}$ and yellow for $\mathbf{L\pi}$. We compare the fraction of eccentric BBHs in different morphologies with their counterparts in circular orbits.  The boundary between different spin morphologies for BBHs in circular orbits has been represented by black dashed lines. We observe that the fraction $f$ of binaries being captured in different morphologies at $a=10\mathbf{M}$ is almost independent of the initial eccentricities.  
In fact, the presence of eccentricity  does not change the response of a population of BBHs to spin precession, and as a result, the number of eccentric BBHs in different morphologies are almost identical to BBHs in circular orbits.

\begin{figure}[!ht]
    \includegraphics[width=\textwidth]{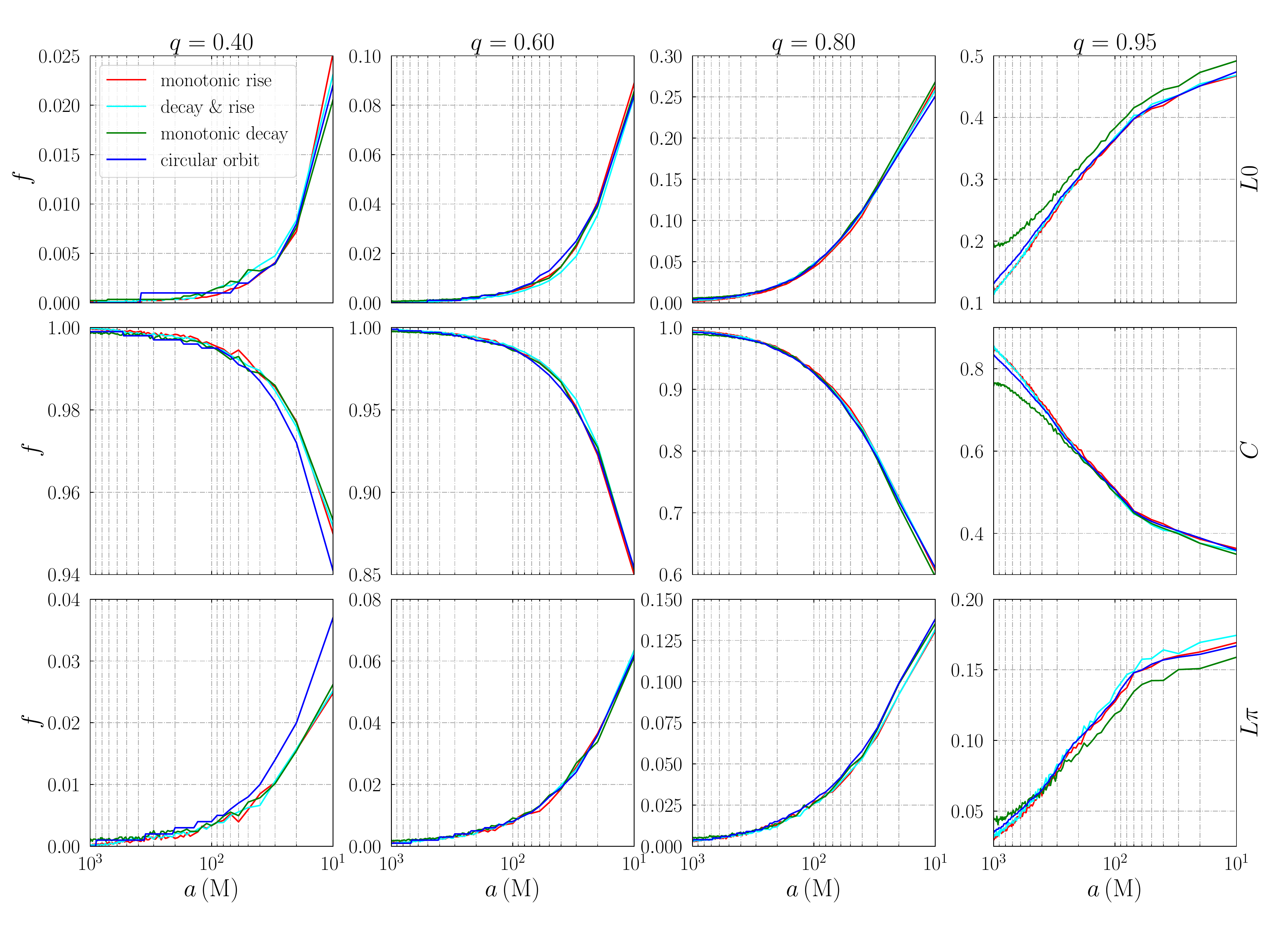}
    \caption{ \label{Fig:morph_evol}Evolution of the fraction of binaries  in $\mathbf{C}$, $\mathbf{L0}$ and $\mathbf{L\pi}$ morphologies during inspiral.  
    The BBHs have spin magnitudes in the range $0.2 \le \chi_i \le 1.0$. The first, second and third rows represent the fraction of binaries $f$ residing in the $\mathbf{L0}$, $\mathbf{C}$ and $\mathbf{L\pi}$ morphologies, respectively, for four mass ratios $q\in \{ 0.40, 0.60, 0.80, 0.95\}$. The red, cyan, green colored lines represent BBHs with eccentricities that rise monotonically, rise after decaying to minimal eccentricities, and monotonically decay, respectively, during the evolution. For comparison, the evolution of $f$ for binaries with zero eccentricity is also plotted in blue.}   
 \end{figure}

We next study the transition of BBHs in eccentric orbit to different morphologies during their inspiral. We compute the number of eccentric BBHs in different morphologies at each orbital separation, and the results are presented in Fig.~\ref{Fig:morph_evol}. As in Figs.~\ref{Fig:morph_mono_increase}, \ref{Fig:morph_decrease_rise} and \ref{Fig:morph_mono_decay}, we evolve eccentric BBHs having three distinct eccentricity evolution patterns: (i) eccentricity rising monotonically (red), (ii) eccentricity rising after decaying to minimal value (cyan), and (iii) eccentricity monotonically decaying (green). For comparison, we also evolve BBHs in quasi-circular orbits as shown in blue. We consider four mass-ratios $q=\{0.4, 0.6, 0.8, 95\}$ for binaries while the spin magnitudes are uniformly distributed in the range $[0.2, 1.0]$.  The first, second and third rows of Fig.~\ref{Fig:morph_evol} show the fraction of binaries residing in $\mathbf{L0}$-morphology,  $\mathbf{C}$-morphology  $\mathbf{L\pi}$-morphology, respectively. At the initial separation, the population of BBHs is dominated by a sample of BBHs in the circulating morphology $\mathbf{C}$-morphology. As expected, the probability of the BBHs to transition to librating morphologies, $\mathbf{L0}$-morphology and $\mathbf{L\pi}$-morphology, increases as binaries inspiral towards merger.  This transition is strongly dependent on mass ratio $q$. With an increase in mass asymmetry, the transition probability towards the librating morphologies decreases as also has been shown in Ref.~\cite{Gerosa:2015tea}. From Fig.~\ref{Fig:morph_evol}, it is evident that eccentricity of BBHs does not bear much influence on morphology transition. We compare the fraction $f$ of binaries in eccentric orbit in all three eccentricity evolution cases in all three morphologies  to the fraction $f$ of circular binaries in their spin morphologies. We see that the pattern of evolution of the number of eccentric BBHs in their respective morphologies is very similar to those in circular orbits.

Past studies have shown the pivotal role of mass ratio $q$ as well as spin magnitudes, $\chi_1$ and $\chi_2$, on binary's precession dynamics~\cite{Kesden:2010yp,Gerosa:2015tea}. The spin-induced growth of eccentricity is a consequence of spin precession that contributes at 2PN order in the long radiation-reaction timescale. Although the 2PN spin-spin interaction terms in Eq.~(\ref{eq:e2}) depend on mass ratio q, they have a negligible effect on the evolution of eccentricities. The spread of eccentricity in Fig.~\ref{Fig:ecc_evol} is mainly attributed to the varying rate of change of eccentricity with $\chi_1$ and $\chi_2$ and associated angular parameters. For given spin magnitudes and orientations, the eccentricity evolution weakly depends on the mass ratio. Since the spin magnitudes affect morphology transition of binaries in circular orbits and also the eccentricity evolution, one would expect that the presence of eccentricity can affect precession dynamics or the probability of binaries being captured in one of the morphologies. However, in a statistical sense, we observe no such dramatic differences due to the presence of eccentricity. In Figs.~\ref{Fig:morph_mono_increase}, \ref{Fig:morph_decrease_rise} and \ref{Fig:morph_mono_decay}, we see small observational differences between the fraction of eccentric BBHs getting captured in the spin morphologies for all three evolutionary patterns of eccentricity and the fraction of circular BBHs in respective morphologies at $a\,=\,10\mathbf{M}$. These differences are within the Poisson counting error except for one case as can be seen in the bottom-middle panel of Fig. \ref{Fig:morph_mono_decay} where the black-dashed line differs from the black-solid line perceptively for $q=0.8$. The similarity between the results for eccentric and circular binaries is consistent throughout the inspiral as can be seen in Fig.~\ref{Fig:morph_evol}. These observations imply that eccentricity has a sub-dominant role in the morphological classification of spin precession, although eccentricity evolution is spin dependent. We further argue that the spin-induced eccentricity growth and morphology transition are disentangled phenomena.

In Ref.~\cite{Kesden:2010yp}, it was noted that the isotropic spin distributions at $a=1000 \mathbf{M}$ remain isotropic at $a=10 \mathbf{M}$ for binaries in quasi-circular orbit. We found similar results for binaries in eccentric orbits considered in this paper.  That is, the isotropic spin distributions at large orbital separation remain isotropic at late stages of inspiral for binaries in eccentric orbits. The individual spin angles $\theta_1,\, \theta_2$ vary on both $\tpre$ and $\trr$, which essentially randomize their distributions over a long timespan and do not provide complete information about binaries' initial spin distribution. Morphology measurement, on the other hand,  provides information about the behavior of spinning binaries on a particular timescale (precession timescale) and is a more robust way to study the spin dynamics of BBHs. The spin morphologies of binaries remain constant on $\tpre$ and evolve slowly on the radiation-reaction time $t_{rr}$. It has been argued that morphology estimation in GW detectors sensitivity window can be used to infer the spin orientations of BBHs at formation, which further be useful in constraining the formation channels~\cite{Gerosa:2017xik,Gerosa:2015tea}. A recent paper~\cite{Gerosa:2018wbw} gave insights into possible correlations between supernovae physics and morphologies of binaries in GW detectors band. The rationale behind tracking previous spin orientations of circular BBHs using morphology estimation is that the binaries from different spin morphologies populate distinct regions in ($\theta_1-\theta_2$) plane of parameter space. The same argument is applicable to BBHs in eccentric orbits. In non-zero eccentricity case, the boundaries separating the morphologies are similar to those separating morphologies of precessing BBHs in circular orbits. Hence, morphology measurement serves as a powerful tool to explore physics of the formation of BBHs in eccentric orbit as well.

\section{Conclusions}\label{sec:conclusion}
Recently, a robust method based on the identification of three mutually exclusive spin morphologies has been developed to describe the dynamics of BBHs in circular orbits \cite{Kesden:2014sla, Gerosa:2015tea}. These morphologies remain constant in a precession cycle while evolving slowly under radiation reaction timescale. In this paper, we apply this spin morphology-based classification of precession dynamics to generic BBHs in eccentric orbits. 
We evolved a population of BBHs from an initial separation $a = 1000\mathbf{M}$ to a = $10\mathbf{M}$ using orbit-averaged precession equations while incorporating higher-order spin-spin contributions in the derivatives of eccentricity and mean motion. We found that the eccentricities of a population of BBHs obey three distinctive evolutionary patterns that depend on their initial eccentricities and BH spin magnitudes. These evolutionary patterns are: (i) eccentricity monotonically increasing until final orbital separation, (ii) eccentricity rising after decaying to a minimum value, and (iii) eccentricity monotonically decreasing throughout the inspiral. The rise in eccentricity to non-negligible values is due to the 2PN order positive gradient induced by the spin-spin contribution in the derivative of eccentricity. 

Depending on the spin orientations ($\theta_1, \theta_2$), spin magnitudes ($\chi_1, \chi_2$) and mass ratio $q$, a BBH falls in a particular spin morphology during inspiral. The BBH can undergo transitions from one morphology to another morphology during its inspiral phase. We studied the effects of the three different evolutionary patterns of eccentricities on the morphology transition of BBH population and compared them with that of BBHs in quasi-circular orbits. We found that eccentricity plays a sub-dominant role on the spin morphology of precessing BBH.  The transition probability of a population of BBHs in eccentric orbits to different morphologies during inspiral is similar to that of BBHs in circular orbits. The statistical independence of morphology transition from eccentricity indicates that the morphology classification of BBHs is also useful for binaries in eccentric orbits and can help probe their formation scenarios as in the case of binaries in circular orbits.

Understanding the formation mechanism of compact binaries is an outstanding problem in astronomy. The compact binaries observable by ground-based detectors are likely to be nearly circular, but the plausibility of observing binaries with small eccentricities cannot be ruled out.  It has been shown that earth-based GW observatories could differentiate between field and cluster formation by looking at spin dynamics, redshift distribution and possibly kicks, while assuming binaries to be circular in the detectors' band. The BBHs formed both in the field, and cluster environments can have measurable eccentricities in the space-based GW detectors like LISA~\cite{2017arXiv170200786A}. For such binaries, mass and eccentricity in LISA band are used to discriminate between different formations channels. Spin measurements of BBHs in the LISA band provide another means to constrain the formation mechanism. In our study, we have shown that eccentricities do not diminish the robustness of spin dynamics in predicting initial spin distributions. In the future, we plan to extend this study by implementing a full treatment of the spin dynamics-eccentricity distribution of the BBHs observable by LISA originating from both dynamical processes in the dense stellar cluster and isolated binary evolution in galactic fields.

\section*{Acknowledgments}
 It is a pleasure to thank Davide Gerosa for encouraging us to pursue this problem and for several discussions on this subject. We also thank him and Richard O'Shaughnessy for carefully reading the manuscript and making useful comments. We thank Antoine Klein, Nathan Johnson-McDaniel, and Ofek Birnholtz for discussions and comments. This work was supported in part by the Navajbai Ratan Tata Trust. A.G. acknowledges support from the NSF grants AST-1716394 and AST-1708146. PJ and KSP acknowledge funding from the Science and Engineering Research Board (SERB), Government of India. We acknowledge the use of IUCAA LDG cluster Sarathi for the numerical work. The figures in this paper are generated using python-matplotlib package~\cite{Hunter:2007}. This paper has been assigned the LIGO preprint number P1900084. 

\appendix

\section{PN equations}
\label{appen:pn_coefficients}
The coefficients of the first-order ordinary differential equations of $n$ at different PN order are expressed as~\cite{Klein:2018ybm, Klein:2010ti, Arun:2009mc},

\begin{subequations}\label{E:RR2}
	\begin{align}
		\dot{n}^{}_{\text{N}} &=\frac{1}{(1-e^2)^{7/2}} \left(\frac{96}{5}+\frac{292}{5}\,e^2+\frac{37}{5}\,e^4\right)\,,\\
        \dot{n}^{}_{\text{1PN}} &= \frac{1}{(1-e^2)^{9/2}}  \left[ -\frac{\numprint{4846}}{35} - \frac{264}{5}\eta + e^2\left(\frac{\numprint{5001}}{35}-570\,\eta \right) + e^4\left( \frac{\numprint{2489}}{4}-\frac{\numprint{5061}}{10}\,\eta\right) + e^6\left(\frac{\numprint{11717}}{280}-\frac{148}{5}\,\eta\right) \right]\,,\\
		\dot{n}^{}_{\text{1.5PN}} &= \frac{384}{5}\,\pi\, \varphi(e)  - \frac{1}{10\left(1 - e^2\right)^{5}} \beta \left( 3088 + \numprint{15528}e^2 + 7026 e^4 + 195e^6,\, 2160 + \numprint{11720} e^2 \right. \nn\\ 
		    &\left.+ 5982 e^4 + 207 e^6 \right) \,,\\
		\dot{n}^{}_{\text{2PN}} &= -\frac{1}{160(1-e^2)^{11/2}}\,\sigma \left( \numprint{21952} + \numprint{128544}e^2 + \numprint{73752}e^4 + \numprint{3084}e^6,\, \numprint{64576} + \numprint{373472}e^2 + \numprint{21021}e^4 + \numprint{8532}e^6,\right.\nn \\
		    & \left. \numprint{131344}e^2 + \numprint{127888}e^4 + \numprint{7593}e^6 \right)  + \frac{1}{320(1-e^2)^{11/2}}\,\tau\left(448 + \numprint{4256}e^2 + \numprint{3864}e^4 + 252e^6,\right.\nn \\
		    &\left. 64 + 608e^2 + 552e^4 + 36e^6,\, 16e^2 + 80e^4 +9e^6\right) + \frac{1}{(1-e^2)^{11/2}} \left[ -\frac{1159}{945} + \frac{\numprint{15265}}{21}\,\eta + \frac{944}{15}\,\eta^2 \right. \nn \\
		    & + e^2\left( -\frac{\numprint{975868}}{189}  + \frac{\numprint{10817}}{5}\,\eta + \frac{\numprint{182387}}{90}\,\eta^2\right) + e^4 \left( -\frac{\numprint{211193}}{90} - \frac{\numprint{955709}}{140}\,\eta + \frac{\numprint{396443}}{72}\,\eta^2 \right) + e^6\left(\frac{\numprint{4751023}}{1680}\right.\nn \\
		    &\left. -\frac{\numprint{203131}}{48}\,\eta + \frac{\numprint{192943}}{90}\,\eta^2\right) + e^8\left(\frac{\numprint{391457}}{3360}-\frac{6037}{56}\eta + \frac{2923}{45}\eta^2\right)+ \sqrt{1-e^2}\left\{ 48-\frac{96}{5}\,\eta + e^2\left( 2134 - \frac{4268}{5}\,\eta \right) \right.\nn \\
		    &\left.\left. + e^4\left(2193 - \frac{4386}{5}\,\eta\right) + e^6\left( \frac{175}{2}-35\,\eta\right) \right\} \right]\,,\\
		\dot{n}^{}_{\text{2.5PN}} &=  \frac{96}{5}\,\pi  \left[ -\frac{\numprint{17599}}{672} \, \psi_n(e) - \frac{189}{8}\eta\, \zeta_n(e) \right] \,,\\
		\dot{n}^{}_{\text{3PN}} &=  \frac{1}{(1-e^2)^{13/2}}\left[     \frac{\numprint{4915859933} }{\numprint{1039500}} +\left( \frac{\numprint{1463719}}{\numprint{2268}} - \frac{369}{10}\pi^2\right)\,\eta - \frac{\numprint{711931}}{420}\,\eta^2 -\frac{\numprint{1121}}{27}\,\eta^3 + e^2  \left(  \frac{\numprint{76740432133}}{\numprint{2079000}} \right. \right. \nn \\
		    &\left.+ \left\{  \frac{\numprint{140649817}}{\numprint{3240}} + \frac{\numprint{24777}}{80}\,\pi^2  \right\}\,\eta - \frac{\numprint{17171137}}{840}\,\eta^2 - \frac{\numprint{1287385}}{324}\,\eta^3 \right)  + e^4 \left( -\frac{\numprint{136998957827}}{\numprint{8316000}}  
		     + \left\{ \frac{\numprint{11146580197}}{\numprint{90720}} \right.\right.\nn \\ 
		    &\left.\left.-\frac{\numprint{26887}}{160}\,\pi^2 \right\}\,\eta  +  \frac{\numprint{18119597}}{3360}\,\eta^2 - \frac{\numprint{33769597}}{1296}\,\eta^3 \right) + e^6 \left(  - \frac{\numprint{88115571763}}{\numprint{5544000}} +\left\{  \frac{\numprint{4653403}}{4032} -\frac{\numprint{59093}}{160}\,\pi^2 \right\}\,\eta \right.\nn \\
		    &+ \frac{\numprint{123833019}}{2240}\,\eta^2 - \frac{\numprint{3200965}}{108}\,\eta^3 \Bigg) + e^8 \left( \frac{\numprint{59327921801}}{7392000} + \left\{  -\frac{\numprint{9471607}}{672}-\frac{\numprint{12177}}{640}\,\pi^2 \right\}\,\eta + \frac{\numprint{2260735}}{168}\,\eta^2\nn \right. \\
		    &\left.-\frac{982645}{162}\,\eta^3 \right) + e^{10} \left(  \frac{\numprint{33332681}}{\numprint{197120}} - \frac{\numprint{1874543}}{\numprint{10080}}\,\eta + \frac{\numprint{109733}}{\numprint{840}}\,\eta^2-\frac{\numprint{8288}}{81}\,\eta^3\right) + \sqrt{1-e^2} \left[ -\frac{\numprint{2667319}}{1125}+ \left\{ \frac{\numprint{56242}}{105} \right.\right. \nn\\
		    &\left.-\frac{41} {10}\,\pi^2\right\}\,\eta + \frac{632}{5}\,\eta^2 + e^2\left( -\frac{\numprint{2673296}}{375} + \left\{ -\frac{\numprint{10074037}}{315} + \frac{\numprint{45961}}{240}\,\pi^2\right\}\,\eta + \frac{\numprint{125278}}{15}\,\eta^2\right) + e^4 \left( \frac{\numprint{700397951}}{\numprint{21000}} \right.\nn \\
		    &\left. + \left\{ -\frac{\numprint{4767517}}{60} + \frac{6191}{32}\,\pi^2\right\}\,\eta + \frac{\numprint{317273}}{15}\,\eta^2 \right) +e^6 \left( \frac{\numprint{708573457}}{\numprint{31500}} +\left\{ -\frac{\numprint{6849319}}{252} + \frac{287}{960}\,\pi^2\right\}\,\eta + \frac{\numprint{232177}}{30}\,\eta^2 \right) \nn \\
		    &\left. +e^8\left( \frac{\numprint{56403}}{112} - \frac{\numprint{427733}}{840}\,\eta + \frac{4739}{30}\,\eta^2 \right)\right]  + \left( \frac{\numprint{54784}}{175} + \frac{\numprint{465664}}{105}\,e^2 + \frac{\numprint{4426376}}{525}\,e^4 + \frac{\numprint{1498856}}{525}e^6 + \frac{\numprint{31779}}{350}e^8\right)\nn \\
		    &\left.\times \log{\left(\frac{x}{x_0} \frac{1+\sqrt{1-e^2}}{2(1-e^2)}\right)} \right] + \frac{96}{5} \left[ -\frac{\numprint{116761}}{3675}\kappa(e) + \left\{ \frac{16}{3}\pi^2 - \frac{1712}{105}C - \frac{1712}{105}\ln(4\pi x^{3/2} r_0)\right\} F(e)\right]\,.
	\end{align}
\end{subequations}
The coefficients of $de^2/dt$ are, 
\begin{subequations}\label{E:RR3}
	\begin{align}
    \dot{e^2}^{}_{\text{N}} &=  \frac{2e^2}{(1-e^2)^{5/2}}\left(\frac{304}{15}+\frac{121}{15}\,e^2\right)\,\\
    \dot{e^2}^{}_{1\text{PN}} &= \frac{2\,e^2}{(1-e^2)^{7/2}} \left[ -\frac{939}{35}-\frac{4084}{45}\,\eta + e^2 \left( \frac{\numprint{29917}}{105}-\frac{7753}{30}\,\eta\right) +e^4\left( \frac{\numprint{13929}}{280}-\frac{1664}{45}\,\eta\right) \right]\,,\\
    %
    %
    \dot{e^2}^{}_{1.5\text{PN}} &= - \frac{e^2}{15(1-e^2)^4}\, \beta\left( \numprint{13048} + \numprint{12000}e^2 + 789e^4,\, \numprint{9208} +\numprint{10026}e^2 + 835e^4  \right)  \nn \\
        &+ \frac{64e^2}{5}\left[  \frac{985}{48}\pi \varphi_e(e) \right]\,,\\
        %
        %
    \dot{e^2}^{}_{2\text{PN}} &= \frac{2e^2}{ (1-e^2)^{9/2}} \left[ -\frac{\numprint{961973}}{1890} + \frac{\numprint{70967}}{210}\,\eta + \frac{752}{5}\,\eta^2 + e^2\left( -\frac{\numprint{3180307}}{2520} -\frac{\numprint{1541059}}{840}\,\eta + \frac{\numprint{64433}}{40}\,\eta^2\right)\right. \nn\\
        & + e^4\left( \frac{\numprint{23222071}}{\numprint{15120}} - \frac{\numprint{13402843}}{5040}\,\eta + \frac{\numprint{127411}}{90}\,\eta^2\right)  + e^6 \left( \frac{\numprint{420727}}{3360}-\frac{\numprint{362071}}{2520}\,\eta + \frac{821}{9}\,\eta^2 \right) \nn \\
        & \left. + \sqrt{1-e^2}\left\{ \frac{1336}{3} - \frac{2672}{15}\,\eta + e^2 \left( \frac{2321}{2}-\frac{2321}{5}\,\eta\right) + e^4\left( \frac{565}{6} - \frac{113}{3}\,\eta \right)\right\}\right] \nn \\
        & - \frac{1}{240(1-e^2)^{9/2}}\, \sigma\left(-320 + \numprint{101664}e^2 + \numprint{116568}e^4 + \numprint{9420}e^6,\, -320 + \numprint{296672}e^2 + \numprint{333624}e^4 \right. \nn \\
        &\left. +\numprint{26820}e^6,\, \numprint{88432}e^2 + \numprint{161872}e^4  + \numprint{16521}e^6\right) + \frac{1}{480(1-e)^{9/2}}\,\tau\left(-320 + \numprint{2720}e^2 + \numprint{5880}e^4 + 540e^6,\right.\nn\\
        &\left. -320 - 160e^2 + \numprint{1560}e^4 + 180e^6, \,  16e^2 + 80e^4 + 9e^6\right) \,,\\
	%
	%
	\dot{e^2}^{}_{2.5\text{PN}} &=  -\frac{64}{5}e^2\,\pi\, \left[  \frac{\numprint{55691}}{\numprint{1344}}\psi_e (e)+ \frac{\numprint{19067}}{126}\,\eta\,\zeta_e(e)\right]\,,\\
	%
	%
	\dot{e^2}^{}_{3\text{PN}} &= \frac{2e^2}{(1-e^2)^{11/2}}\left[ \frac{\numprint{54177075619}}{\numprint{6237000}} + \left( \frac{\numprint{7198067}}{\numprint{22680}}+ \frac{\numprint{1283}}{10}\,\pi^2\right)\,\eta-\frac{\numprint{3000281}}{\numprint{2520}}\,\eta^2-\frac{\numprint{61001}}{486}\,\eta^3 + e^2\left(\frac{\numprint{6346360709}}{891000} \right.\right.\nn \\
		&\left. + \left\{  \frac{\numprint{9569213}}{360}+\frac{ \numprint{54001}}{960}\,\pi^2\right\}\,\eta  + \frac{\numprint{12478601}}{\numprint{15120}}\,\eta^2 -\frac{\numprint{86910509}}{\numprint{19440}}\,\eta^3 \right)  + e^4\left( -\frac{\numprint{126288160777}}{\numprint{16632000}} + \left\{ \frac{\numprint{418129451}}{\numprint{181440}} \right. \right.\nn \\
		&\left. \left. -\frac{\numprint{254903}}{\numprint{1920}}\,\pi^2\right\}\,\eta+  \frac{\numprint{478808759}}{\numprint{20160}}\,\eta^2 - \frac{\numprint{2223241}}{180}\,\eta^3 \right) + e^6 \left( \frac{\numprint{5845342193}}{\numprint{1232000}} + \left\{ -\frac{\numprint{98425673}}{\numprint{10080}} -\frac{\numprint{6519}}{\numprint{640}} \pi^2 \right\}\,\eta \nn \right. \\
		&\left. +\frac{\numprint{6538757}}{630}\,\eta^2 - \frac{\numprint{11792069}}{\numprint{2430}}\,\eta^3\right) + e^8 \left( \frac{\numprint{302322169}}{\numprint{1774080}} -\frac{\numprint{1921387}}{\numprint{10080}}\,\eta + \frac{\numprint{41179}}{216}\,\eta^2 - \frac{\numprint{193396}}{1215}\,\eta^3 \right)  \nn \\
		&  +\sqrt{(1-e^2)} \left[  -\frac{\numprint{22713049}}{\numprint{15750}} + \left\{ -\frac{\numprint{5526991}}{945} + \frac{8323}{180}\,\pi^2 \right\}\,\eta + \frac{54332} {45}\,\eta^2 + e^2 \left( \frac{\numprint{89395687}}{\numprint{7875}} +\left\{-\frac{\numprint{38295557}}{1260}\right.\right.\right. \nn \\
		&\left. \left. +\, \frac{\numprint{94177}}{960} \,\pi^2\right\}\,\eta + \frac{\numprint{681989}}{90}\,\eta^2 \right)  + e^4 \left( \frac{5321445613}{378000} + \left\{ -\frac{\numprint{26478311}}{1512} + \frac{2501}{2880}\,\pi^2\right\}\,\eta + \frac{\numprint{225106}}{45}\eta^2 \right)  \nn \\
		&\left. +\,e^6\left( \frac{\numprint{186961}}{336} - \frac{\numprint{289691}}{504}\,\eta + \frac{3197}{18}\eta^2 \right) \right] + \frac{\numprint{730168}}{\numprint{23625}} \frac{1}{1+\sqrt{1-e^2}} \nn \\
		&\left. + \frac{304}{15}\left( \frac{\numprint{82283}}{1995} + \frac{\numprint{297674}}{1995}e^2 + \frac{\numprint{1147147}}{\numprint{15960}}e^4 + \frac{\numprint{61311}}{\numprint{21280}}e^6\right) \log{\left(\frac{x}{x_0} \frac{1+\sqrt{1-e^2}}{2(1-e^2)} \right)}\right] -\frac{64}{5}\,e^2\,\left[\left\{ \frac{\numprint{89789209}}{\numprint{352800}} \right.\right. \nn \\
		&\left.\left. - \frac{\numprint{87419}}{630}\ln 2 + \frac{\numprint{78003}}{560} \ln 3 \right\} \kappa_e(e) - \frac{769}{96} \left( \frac{16}{3} \pi^2 - \frac{1712}{105} \left \{ C + \ln \left(4x^{3/2} r_0 \right) \right \}    \right) F_e(e) \right]\,,
		  \end{align}
\end{subequations}

where, 
\begin{subequations}
\label{eq:beta:sigma:tau}
\begin{eqnarray}
   \beta(a,b) &=& \hat{\mathbf{J}}\cdot\left[ a\,\left( \Sa + \Sb\right)  + b\, \left( q\Sa + q^{-1}\Sb \right) \right]\,,\\
   \sigma(a,b,c) &=& \frac{1}{\eta} \left[ a\, \Sa\cdot\Sb -b\,(\hat{\mathbf{J}}\cdot\Sa) (\hat{\mathbf{J}}\cdot\Sb) + c\,\mid\hat{\mathbf{J}} \times \Sa\mid \mid\hat{\mathbf{J}} \times \Sb\mid\,\cos( \psi_1 + \psi_2)  \right]\,,\\
   \tau(a,b,c) &=& \sum_{i=1}^{2}\frac{1}{m_i^2}\left[a\,\mathbf{S}^2_i -b\,(\hat{\mathbf{J}}\cdot\mathbf{S}_i)^2 + c\,\mid\hat{\mathbf{J}}\times \mathbf{S}_i \mid^2 \cos2\psi_i\right]\,.
\end{eqnarray}
\end{subequations}
The angles $\psi_i\, (i = 1,2)$ are subtended by the line of periastron and the projections of spins on the orbital plane, as shown in Fig.~\ref{fig:ref-frame}. The enhancement factors appearing in the above equations are expressed as follows,

\begin{subequations}
\begin{eqnarray}
\varphi(e)&=& 1 + \frac{\numprint{2335}}{192}e^2 + \frac{\numprint{42955}}{768}e^4 \,,\\ 
\psi_{n}(e) &=& \frac{\numprint{1344}}{\numprint{17599}}\frac{7-5e^2}{1-e^2} \varphi(e) + \frac{\numprint{8191}}{\numprint{17599}}\psi(e)\,,\\
\zeta_{n}(e) &=& \frac{583}{567}\zeta(e) - \frac{16}{567}\varphi(e) \,,\\
\zeta(e) &=& 1 + \frac{\numprint{1011565}}{\numprint{48972}} e^2 + \frac{\numprint{106573021}}{\numprint{783552}}e^4\,,\\
\psi(e) &=& 1- \frac{\numprint{22988}}{\numprint{8191}}e^2- \frac{\numprint{36508643}}{\numprint{524224}}e^4\,,\\
\kappa(e) &=& 1+ \left( \frac{62}{3} - \frac{\numprint{4613840}}{\numprint{350283}}\ln2 + \frac{\numprint{24570945}}{\numprint{1868176}}\ln3 \right)e^2 + \left( \frac{9177}{64}+\frac{\numprint{271636085}}{\numprint{1401132}}\ln 2 - \frac{\numprint{466847955}}{\numprint{7472704}}\ln 3\right)e^4 \,,\\
F(e)&=& 1 + \frac{62}{3}e^2 + \frac{\numprint{9177}}{64}e^4\,,\\
\varphi_e(e) &=& \frac{192}{985}\frac{\sqrt{1-e^2}}{e^2}\left[\sqrt{1-e^2}\,\varphi(e)-\Tilde{\varphi}(e)\right]\,,\\
\Tilde{\varphi}(e)&=&1+\frac{209}{32}e^2 + \frac{2415}{128}e^4 \,,\\
\psi_e(e) &=&  \frac{\numprint{18816}}{\numprint{55691}} \frac{1}{e^2\sqrt{1-e^2}}\left[ \sqrt{1-e^2}\left(1-\frac{11}{7}e^2\right)\varphi(e)-\left(1-\frac{3}{7}e^2\right)\Tilde{\varphi}(e)\right] \nn \\&&+\, \frac{\numprint{16382}}{\numprint{55691}}\frac{\sqrt{1-e^2}}{e^2} \left[ \sqrt{1-e^2}\psi(e)-\Tilde{\psi}(e)\right] \,,\\
\Tilde{\psi}(e) &=& 1-\frac{\numprint{17416}}{8191}e^2 -\frac{\numprint{14199197}}{\numprint{524224}}e^4\,,\\
\zeta_e (e) &=& \frac{924}{\numprint{19067}}\frac{1}{e^2\sqrt{1-e^2}}\left[-(1-e^2)^{3/2}\varphi(e) + \left(1-\frac{5}{11}e^2\right)\Tilde{\varphi}(e) \right] + \frac{\numprint{12243}}{\numprint{76268}}\frac{\sqrt{1-e^2}}{e^2}\left[\sqrt{1-e^2}\zeta(e)-\Tilde{\zeta}(e)\right] \,,\\
\Tilde{\zeta} (e) &=& 1+ \frac{\numprint{102371}}{8162}e^2 + \frac{\numprint{14250725}}{\numprint{261184}}e^4\,,\\
\kappa_e (e) &=& \frac{\sqrt{1-e^2}}{e^2}\left[\sqrt{1-e^2}\kappa(e) -\Tilde{\kappa}(e)\right]\left(\frac{769}{96}-\frac{\numprint{3059665}}{\numprint{700566}}\ln 2 + \frac{\numprint{8190315}}{\numprint{1868176}}\ln 3\right)^{-1}\,,\\
\Tilde{\kappa}(e) &=& 1+ \left( \frac{389}{32} - \frac{\numprint{2056005}}{\numprint{233522}}\ln 2 + \frac{\numprint{8190315}}{\numprint{934088}}\ln 3\right)e^2 + \left(\frac{3577}{64} + \frac{\numprint{50149295}}{\numprint{467044}}\ln 2 - \frac{\numprint{155615985}}{\numprint{3736352}}\ln 3 \right)e^4\,,\\
F_e (e) &=& (1-e^2)^{-11/2}\left( 1 + \frac{2782}{769}e^2 + \frac{\numprint{10721}}{6152}e^4 + \frac{1719}{\numprint{24608}}e^6 \right)\,,
\end{eqnarray}
\end{subequations}

where $C = 0.577$ is Euler's Constant, $x_0 =1$ and $r_0 = 1$ are scaling parameters. 

\bibliographystyle{apsrev4-1}
\bibliography{sor_main.bib}

\end{document}